\documentclass[shortnote,twocolumn]{jpsj2}
\usepackage{graphicx}

\def\runtitle{a sample}
\def\runauthor{Andrzej \textsc{Eilmes} and Rudolf A. \textsc{R\"{o}mer}}

\title{%
  Divergences of the localization lengths in the two-dimensional,
  off-diagonal Anderson model on bipartite lattices}

\author{%
  Andrzej \textsc{Eilmes}$^{1}$ and Rudolf A.
  \textsc{R\"{o}mer}$^{2,}$\thanks{Permanent address: Physics
    Department, University of Warwick, Coventry CV4 7AL, UK,
    Email: r.roemer@warwick.ac.uk} }

\inst{%
$^1$Faculty of Chemistry, Jagiellonian University, Ingardena 3, 30-060 Krak\'{o}w, Poland \\
$^2$Institut f\"{u}r Physik, Technische Universit\"{a}t, D-09107 Chemnitz,
  Germany
}

\recdate{$Revision: 1.4 $, compiled \today}

\abst
{
}

\kword{%
localization, off-diagonal disorder, critical exponents, bipartiteness
}

\begin{document}
\sloppy
\maketitle

\section{Introduction}

The disorder-induced metal-insulator transition (MIT) and the concept
of Anderson localization \cite{And58,EcoC70,EcoC72,Eco72,LicE74} has
been the subject of intense research activities for more than forty
years.  The scaling theory of localization \cite{AbrALR79} predicts
that for generic, interaction-free situations in 2D all states are
localized; there is no disorder-driven MIT \cite{LeeR85,KraM93,BelK94}
and the system remains an insulator.  However, non-localized states
were found at the band center of an Anderson model with purely
off-diagonal disorder \cite{Oda80,SouE81, FerAE81,PurO81,SouWGE82}.
Numercal studies of such systems revealed that the localization length
diverges at the energy $E=0$ \cite{EilRS98a,EilRS98b}. Scaling
properties of this divergence suggest that the states at the band
center are critical \cite{EilRS98b,XioE01} and that are indeed no
truly extended states in 2D even in this off-diagonal case.

Of special importance is the model on a bipartite lattice with an even
number of sites where the energy spectrum is strictly symmetric around
$E=0$.  In this case, related to the chiral universality class, $E=0$
states are non-localized in any dimension \cite{Weg79,GadW91,Gad93}.
This has been recently demonstrated analytically in 2D and 3D using a
renormalization group (RG) approach \cite{FabC00}.  In Ref.\ 
\cite{FabC00} it has been also suggested that for sufficiently large
energies the divergence of the localization length at the band center
may be described by a power law,
\begin{equation}\label{eq-power-law}
   \xi(E) \propto \left| \frac{E_0}{E} \right|^{\nu}
\end{equation}
whereas it takes more complicated form
\begin{equation}\label{eq-exp-law}
  \xi(E) \propto \exp\sqrt{\frac{\ln E_0/E}{A}}
\end{equation}
below a certain crossover energy $E^{*}$.

Recently \cite{EilRS01} we studied such 2D bipartite systems with
various types of off-diagonal disorder by means of the transfer-matrix
method (TMM) and investigated the divergence of the localization
length close to $E=0$. We showed that the data fit the power-low
behaviour down to energy $E=2 \times 10^{-5}$ and the corresponding
critical exponents seem to depend on the type of disorder distribution
and strength.  Here we extend these results to even smaller energies
and examine honeycomb lattices where the Van Hove singularity at $E=0$
does not interfere with the divergence due to the bipartiteness.

\section{Computation of the localization lenghts and the critical exponents}

We considered a single electron on a 2D lattce described by the
Hamiltonian
\begin{equation}
  H = \sum_{i \neq j}^N t_{ij} \left| i \right\rangle \left\langle j
  \right| 
\label{hamilt}
\end{equation}
where $\left| i \right\rangle$ denotes the electron at site $i$. The
off-diagonal disorder was introduced by {\em random} hopping elements
$t_{ij}$ between nearest neighbor sites. We considered bipartite
square and honeycomb lattices. The latter is topologically equivalent
to a brick-layer structure \cite{SchO91}, which implies that one of
the connections to the nearest-neighbor is absent --- the
corresponding hopping element is equal to $0$.

On both lattices we examined three different distributions of the
off-diagonal elements $t_{ij}$:
\begin{enumerate}
\item a rectangular distribution \cite{EilRS98a}
\begin{displaymath}
P(t_{ij}) =
\left\{ \begin{array}{ll}
    1/w & \textrm{if $\left|t_{ij}-c\right| \leq w/2$,}\\
    0   & \textrm{otherwise,}
\end{array}\right.
\end{displaymath}
\item a Gaussian distribution
\begin{displaymath}
 P(t_{ij})= \frac{1}{\sqrt{2\pi\sigma^{2}}}
            \exp \left[ -\frac{(t_{ij}-c)^2}{2\sigma
    ^2} \right],
\end{displaymath}
\item a rectangular
distribution of the logarithm of $t_{ij}$ \cite{SouWGE82}
\begin{displaymath}
P(\ln t_{ij}/t_0) = \left\{ \begin{array}{ll}
       1/w & \textrm{if $\left| \ln t_{ij}/t_0 \right| \leq w/2$,}\\
       0   & \textrm{otherwise.}
        \end{array} \right.
\end{displaymath}
\end{enumerate}
In the case of rectangular and Gaussian distributions of hopping
elements we have set the width $w$ and the standard deviation $\sigma$
of the distribution to $1$ and centered it at $c=0$. In the case of
logarithmic-$t$ distribution we chose $t_0 = 1$ which set the energy
scale and performed the calculations for two values of the
distribution width, e.g., $w=2$ and $6$.

We used the transfer-matrix method \cite{MacK81,MacK83} to compute the
localization lengths for strips of widths $M$ up to $220$ in the
energy interval $10^{-8} \leq E \leq 0.1$, the actual values depend on
the disorder parameters. The accuracy of the results was $1\%$.  Next,
finite-size scaling \cite{MacK83} was applied to the data.  We used
$M$ as large as possible in order to avoid problems related to the
finite-size effects. For some cases this imposes the restriction on
the smallest energies which we were able to use in FSS.

\begin{figure}[h]
\centerline{\includegraphics[scale=0.3]{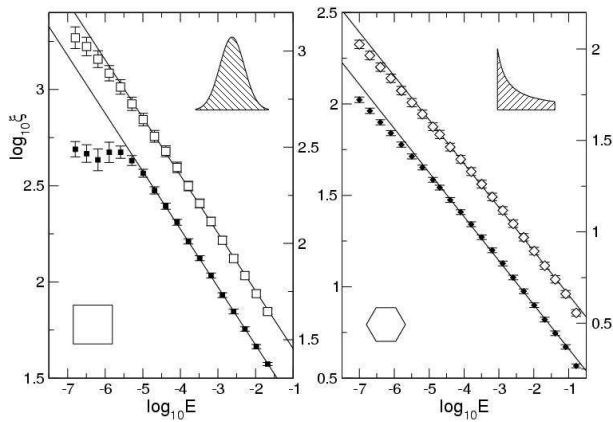}}
\caption{Scaling parameter $\xi$ vs. energy. Left panel: square lattice,
  Gaussian distribution ($\sigma=1$, $c=0$), right panel: honeycomb 
  lattice, logarithmic $t$ distribution ($w=6$); filled symbols: 
  results for $M=50$--$100$, open symbols: results for $M=110$--$160$ 
  (Gaussian distribution) or $M=100$--$150$ (logarithmic distribution).  }
\label{fig-fss-change}
\end{figure}

To investigate the divergence of the localization lenghts at the band
center we plotted the scaling parameter $\xi$ in a double-logarithmic
plot. The deviation of the divergence from the power-law behaviour
should then be easily seen. In most cases we observed that at the
energies close to $E=0$ the divergence is slower than described by a
power-law.  However, these deviations get smaller when we increase
widths $M$ of the systems used in FSS (an example of such behaviour is
shown in Fig. \ref{fig-fss-change}), therefore we relate them 
to finite-size effects.

\section{Results and disussions}

In Table \ref{tab-res} we collected the critical exponents obtained
for different lattices and disorder distributions.  All values are in
the range of $0.2-0.6$ and depend on the disorder type. Generally the
exponents are larger for weaker disorders (i.e.\ when the localization
lengths are larger). The differences between square and honeycomb
lattice in most cases do not exceed the accuracy of the results. The
exception is the logarithmic $t$ distribution with $w=2$, when the
exponent is almost $2$ times larger on a honeycomb lattice.  The large
change in the exponent may be related to the low density of states on
a honeycomb lattice, where the van Hove singularity in the density of
states is absent.

\begin{table}
\caption{\label{tab-res}
  Estimated values of the exponents $\nu$ of the localization lengths
  for various disorder strengths and distributions. The range of the
  strip widths $M$ is displayed in the second column; the power-law
  behaviour is obeyed in the energy ranging from $E_{min}$ up to $1
  \times 10^{-2}$.  The error bars represent the standard deviations
  from the power-law fit and should be increased up to one order of
  magnitude for a reliable representation of the actual errors.}
\begin{tabular}{@{\hspace{\tabcolsep}\extracolsep{\fill}}lccc} 
  \multicolumn{3}{c}  {square lattice}
\\
\hline
disorder  & $M$  & $E_{\rm min}$ & $\nu$ \\ 
\hline
box, $c=0$      & $150-220$ & $8 \times 10^{-7}$ & $0.317 \pm 0.007$  \\
Gaussian, $c=0$ & $110-160$ & $2 \times 10^{-6}$ & $0.303 \pm 0.006$  \\
lnt, $w=2$      & $110-190$ & $2 \times 10^{-6}$ & $0.357 \pm 0.009$  \\
lnt, $w=6$      & $120-170$ & $1 \times 10^{-5}$ & $0.232 \pm 0.007$  \\
\hline
\\
  \multicolumn{3}{c} {honeycomb lattice} \\
\hline
disorder  & $M$  & $E_{\rm min}$ &  $\nu$ \\
\hline
box, $c=0$      & $110-170$ & $2 \times 10^{-8}$ & $0.290 \pm 0.004$ \\
Gaussian, $c=0$ & $120-170$ & $1 \times 10^{-6}$ & $0.273 \pm 0.005$ \\
lnt, $w=2$      & $100-170$ & $2 \times 10^{-4}$ & $0.604 \pm 0.015$ \\
lnt, $w=6$      & $120-170$ & $1 \times 10^{-5}$ & $0.238 \pm 0.007$ \\
\end{tabular}
\end{table}

The localization lengths exhibit power-law behaviour in a wide energy
range, the lower lower bound $E_{\rm min}$ of which is indicated in
Table \ref{tab-res}.  For smaller energies we observe some deviations,
however, these become smaller for larger system widths $M$ we would
argue that these are finite-size effects that will eventually vanish
if one uses large enough system sizes. The exception to this behaviour
is the logarithmic-$t$ disribution for $w=6$, where below an energy $E
\approx 10^{-5}$ the deviation from power law seems to be size
independent. This may be an indication of the crossover to the
non-power-law behavior as predicted in Ref.\ \cite{FabC00} or to a
power-law with a different exponent. However, there is also a
possibilty that this may be an effect of numerical problems which
appear for strong logarithmic disorders.


\end{document}